\documentclass{aa}

\usepackage[varg]{txfonts}

\usepackage{multirow}
\usepackage{graphicx}

\usepackage{booktabs}

\def\ms{\mbox{$M_{\star}$}~}
\def\mh{\mbox{$M_{\rm h}$}}   
\def\mstar {h^{-1}~\rm M_{\sun}}
\def\mhalo {h^{-1}~\rm M_{\sun}}
\def\msun {M$_{\sun}$}

\def\mpc {h^{-1}~\rm Mpc}
\def\sag {{\sc mdpl2-sag}}

\usepackage[normalem]{ulem} 
\usepackage{placeins}
\usepackage[colorlinks, citecolor=blue]{hyperref}
\bibpunct[; ]{(}{)}{;}{a}{}{;}

\begin{document}

\title{Assessing the connection between galactic conformity and assembly-type bias}

%\titlerunning{GC-AB}

\author{Ivan Lacerna\inst{\ref{uda}} \and Nelson Padilla\inst{\ref{Iate},\ref{OAC}} \and Daniela Palma\inst{\ref{uda}}
}  

\institute{
Instituto de Astronom\'ia y Ciencias Planetarias, Universidad de Atacama, Copayapu 485, Copiap\'o, Chile\\
\email{ivan.lacerna@uda.cl} \label{uda} 
\and
CONICET. Instituto de Astronomía Teórica y Experimental (IATE). Laprida 854, Córdoba X5000BGR, Argentina \label{Iate}
\and 
Universidad Nacional de Córdoba (UNC). Observatorio Astronómico de Córdoba (OAC). Laprida 854, Córdoba X5000BGR, Argentina\label{OAC}
}

\abstract
%Context
{
Galaxies in the Universe show a conformity in the fraction of quenched galaxies out to large distances, being much larger
around quenched central galaxies than for star-forming ones.  On the other hand, simulations have shown that the clustering of halos and the galaxies within them depends on secondary properties other than halo mass, a phenomenon termed assembly bias.
}
%Aims
{
Our aim is to study whether samples that show galactic conformity also show assembly bias and to see if the amplitude of these two effects is correlated.
}
%Methods
{ 
We use synthetic galaxies at $z = 0$ from the semi-analytical model  {\sc sag} run on the MultiDark Planck 2 (\textsc{mdpl2}) cosmological simulation and measure both conformity and galaxy assembly bias for different samples of central galaxies at fixed host halo mass.
We focus on central galaxies hosted by low-mass halos of 10$^{11.6}$ $\leq$ \mh/$\mhalo$ $<$ 10$^{11.8}$ because it is a mass range where the assembly bias has been reported to be strong. The samples of central galaxies are separated according to their specific star formation rate and stellar age.
}
%Results
{
We find that the level of conformity shown by our different samples is correlated with the level of assembly bias measured for them.  We also find that removing central galaxies around massive halos
diminishes the conformity signal and lowers the amount of assembly bias.
}
%Conclusions
{
The high correlation in the amplitude of conformity and assembly bias for different samples with and without removing galaxies near massive halos clearly indicates
the strong relationship between both phenomena. 
}

\keywords{}

\maketitle          

\section{Introduction} 
\label{S1}

As part of the current effort to understand galaxy evolution, the
environment in which galaxies reside is still a subject of intense study.  
The reason is that fundamental galaxy properties such as 
star formation rate and color can be affected by the environment, especially for low-mass galaxies \citep[e.g.,][]{Peng+2010full, Bluck+2014, AF+2018}. 
Furthermore, low-mass galaxies show environmental effects on megaparsec scales, typically well beyond the virial radius of galaxy groups and clusters
\citep[e.g.,][]{Wetzel+2012,Bahe+2013,Benitez-Llambay+2013,Cybulski+2014,Campbell+2015,Hearin+2015,Bahe+2017,Goddard+2017_465env,ZhengZheng+2017,Zinger+2018,Duckworth+2019,Kraljic+2019,Pallero+2019,Tremmel+2019,ZhengZheng+2019,Pandey&Sarkar2020,ZhangY+2021,Lacerna+2022}.

An example of the impact of the large-scale environment is the two-halo galactic conformity 
\citep[e.g.,][]{Kauffmann+2013,  Hearin+2015, Hearin+2016, Kauffmann2015, Paranjape+2015, Bray+2016, Berti+2017, Sin+2017, Sin+2019, Calderon+2018, Lacerna+2018MNRAS, Lacerna+2022, RafieferantsoaDave2018, Sun+2018, Tinker+2018, Treyer+2018, ZuMandelbaum2018, Alam+2020, LiL+2021, Ayromlou+2023, OlsenGawiser2023, WangK+2023, Palma_Lacerna+2025}. 
This term is used to describe the correlation between color or star formation activity in low-mass central galaxies, with stellar masses \ms\ < $10^{10.5}$ \msun,  
and their neighbor galaxies in adjacent halos at separations of several megaparsecs. 
\citet{Kauffmann+2013} found an observational two-halo conformity effect between low-mass central\footnote{They used galaxies without relatively bright neighbors as central galaxies.} galaxies with low specific star formation rate (sSFR) or gas content and neighbor galaxies with low sSFR out to scales of 4 Mpc at $z$ $<$ 0.03. %
Cosmological hydrodynamical simulations 
\citep[e.g.,][]{Bray+2016, WangK+2023}, 
semi-analytic models of galaxy formation \citep[e.g.,][]{Lacerna+2018MNRAS, Lacerna+2022, Ayromlou+2023}, and semi-empirical models 
\citep[e.g.,][]{Sin+2017, Tinker+2018} 
also show a correlation in color or star formation between central galaxies and neighboring galaxies at Mpc scales.

Using mock galaxy catalogs, \citet{Sin+2017} found that the two-halo conformity out to projected distances of 3--4 Mpc from central galaxies primarily relates
to the environmental influence of very large neighboring halos. \citet{Lacerna+2022} demonstrated that the low-mass (\ms $\leq$ 10$^{10}$ $\mstar$) central galaxies that lie in the vicinity of massive systems are the ones most responsible for producing
galactic conformity at Mpc scales. They used two galaxy catalogs from different versions of the semi-analytic model {\sc sag} \citep{Cora+2018} applied to the MultiDark Planck 2 cosmological simulation (\citealt{Klypin+2016, Knebe+2018}) 
and the {\sc IllustrisTNG300} cosmological hydrodynamical simulation \citep{Naiman+2018, Nelson+2018, Marinacci+2018, Pillepich+2018_473, Springel+2018, Nelson+2019}, 
and consistently found that central galaxies 
in the vicinity of galaxy groups are primarily responsible for the correlation between the low-mass centrals and neighboring galaxies at large
separations of several megaparsecs.
\citet{Ayromlou+2023} used a galaxy catalog from a semi-analytic model (LGal-A21, \citealt{Ayromlou+2021}) and showed that the conformity signal, but not all of it, arises from central galaxies near massive systems. 

On the other hand, \citet{WangK+2023} found that low-mass central galaxies from the  Sloan Digital Sky Survey (SDSS, \citealt{York+2000, Blanton+2005}) are more quenched in high-density regions, which generally contain a massive halo that would lie close to these low-mass galaxies.  
They also found a similar trend in {\sc IllustrisTNG300}, which, according to the authors, can be entirely explained by ``backsplash''\footnote{\citet{WangK+2023} used the definition of main progenitors of central galaxies at present that were satellites of other halos for two successive snapshots.} galaxies. 
Recently, \cite{Palma_Lacerna+2025} studied the evolution of low-mass central galaxies with \ms\ = $10^{9.5} - 10^{10}$ $h^{-1}$ M$_{\odot}$ near massive groups and clusters of galaxies using the {\sc IllustrisTNG300} and \sag\ catalogs. 
They found that former satellites, that is, central galaxies at present that were satellites in the past, a simple way to define backsplash or fly-by galaxies, play an important role in the two-halo conformity signal in {\sc IllustrisTNG300}. These galaxies can explain the whole signal at $z \sim 1$, while they contribute up to 75 -- 85\% at $z$ $\lesssim$ 0.3. They found a negligible contribution of former satellites to the conformity signal in \sag. Regardless of the specific contribution of former satellites, the results from these works are consistent in that 
the two-halo galactic conformity is primarily produced by low-mass central galaxies near massive halos with halo mass \mh\ $\geq$ 10$^{13}$ M$_{\odot}$. 

Galaxy assembly bias is another phenomenon where the properties of a galaxy affect statistics on large separations well into the 2-halo scale regime.  
It is not unreasonable then to study both effects together, galaxy assembly bias and conformity.
Galaxy assembly bias is often used to refer to the dependence of large-scale galaxy clustering on secondary halo properties beyond the halo mass. In the context of halo occupation distribution (HOD) modeling, galaxy assembly bias is described as the combination of assembly bias and the so-called occupancy variations, i.e., the dependencies of the galaxy content of halos on secondary halo properties at fixed halo mass \citep{Artale+2018, Zehavi+2018, Contreras+2019}. 
Cosmological numerical simulations (semi-analytical models and hydrodynamical models)
tend to show the existence of an assembly-type bias on synthetic galaxies \citep[e.g.,][]{Croton+2007,LacernaPadilla2011,WangL+2013MNRAS.431..600W,Salcedo+2018, Montero-Dorta2020}. However, it is still unclear whether this effect is present in observations 
\citep[e.g.,][]{Skibba+2006, Cooper+2010, WangL+2013MNRAS.433..515W, Lacerna+2014,LinYT+2016, LinYT+2022, Dvornik+2017, Oyarzun+2022,Oyarzun+2024,Ortega-Martinez+2025}. See the review by \cite{WechslerTinker2018} on this topic.

Several works have suggested a relationship between the two-halo galactic conformity with the galaxy assembly bias as a potential manifestation of the large-scale environment on galaxies
\citep[e.g.,][]{Hearin+2015, Hearin+2016, Berti+2017, Paranjape+2015, Lacerna+2018MNRAS, MansfieldKravtsov2020,  Montero-Dorta2020, Hadzhiyska+2023MNRAS.524.2507H, Damsted+2024, Garcia-Quintero+2025, McConachie+2025}.
By using mock catalogs from the Bolshoi simulation, \cite{Hearin+2015} scrambled the star formation rate (SFR) of satellite galaxies, which only marginally influenced the two-halo conformity signal on central galaxies out to $\sim$ 5 $\mpc$. They concluded that the conformity signal is driven by  
galaxy assembly bias on central galaxies. In contrast, \citet{ZuMandelbaum2018} used a “tunable” HOD to show that galactic conformity can be naturally
explained by the combination of halo quenching and the variation of the halo mass function with the environment, without the need for any galaxy assembly bias. \citet{Paranjape+2015} argued from extended HOD models that, only at separations larger than 8 Mpc, there is genuine two-halo conformity driven by the assembly bias of small host halos. 

One would be tempted to argue that since the conformity signal is typically limited to smaller scales,  it would not correspond to 
assembly bias in principle. However, 
\citet{Hadzhiyska+2023MNRAS.524.2507H} conjectured that the galaxy formation process of Emission Line Galaxies (ELGs)  may be dependent on the presence of nearby massive clusters, which in turn is connected with the two-halo conformity. They used ELGs extracted from the MillenniumTNG cosmological hydrodynamical simulation. It is possible that the level of assembly bias shown by ELGs in this simulation 
may be influenced by the conformity produced by nearby massive halos.

An explicit effort to assess the actual level of equivalence between galactic (or galaxy) conformity and assembly bias needs to be made since environmental effects in the vicinity of galaxy groups and clusters could be the cause of these two apparently different phenomena. For instance, \citet{LacernaPadilla2011} claimed that assembly bias of low-mass systems can be explained by old, small structures located near massive halos that are typically at distances of a few megaparsecs. These massive halos could disrupt the growth of near-small objects with some mechanism \citep[e.g., tidal stripping,][]{Hahn+2009} and, therefore, affect their properties, such as halo mass and formation time. There are similar concepts, such as ``arrested development'' \citep{Dalal+2008,Salcedo+2018,SmithW+2024} or ``neighbor bias'', in which the mean value of halo properties depends on, besides the halo mass, the distance to a massive neighbor and the ratio of that neighbor's mass to the halo mass \citep{Salcedo+2018}.

This work aims to make such an effort and explicitly look at the level of equivalence between conformity and assembly bias at large scales. For this, we use the \sag\ catalog to select central galaxies near massive halos that mostly produce the two-halo conformity in the low-mass regime, and study their direct contribution to the assembly-type bias signal. To avoid confusion with other definitions, we use the term assembly-type bias  here, namely the secondary halo bias reflected in the clustering of central galaxies. 
The paper is organized as follows. Section \ref{S2} describes the 
synthetic galaxy catalog used in this work.
The analysis of the two-halo galactic conformity is shown in Section \ref{S_GC}, whereas the results on the assembly-type bias are shown in Section \ref{S_AB}. We discuss our results and present our conclusions 
in Section \ref{Sconclusions}.

Throughout this paper, the reduced Hubble constant, $h$, is
defined as $H_0 = 100$ $h$ km s$^{-1}$ Mpc$^{-1}$.
We opted for scaling $h$ explicitly throughout this paper with the following dependencies unless the value of $h$ is specified:
 stellar mass and halo mass in $\mstar$, physical scale in $\mpc$,  and sSFR
 in $h$ yr$^{-1}$.

%%%%%%%%%%%%%%%%%%%%%%%%%%%%%%%%%
\section{The \sag\ galaxy catalog} 
\label{S2}
%%%%%%%%%%%%%%%%%%%%%%%%%%%%%%%%%

We use the \sag\ galaxy catalog constructed by combining the semi-analytic model of galaxy formation {\sc sag} \citep{Cora+2018} with the  dark matter only MultiDark Planck 2 (\textsc{mdpl2}) cosmological simulation (\citealt{Klypin+2016, Knebe+2018}). The {\sc sag} code includes the contribution of several physical processes, such as radiative gas cooling, quiescent star formation, starbursts triggered by mergers and disc instabilities, active galactic nuclei (AGN) and supernovae (SNe) Type Ia and II feedback, and chemical enrichment.
The dark matter (DM) simulation assumes a $\Lambda$CDM cosmology with $\Omega_\textrm{m} = 0.307$, $\Omega_{\Lambda} = 0.693$, $\Omega_\textrm{B} = 0.048$, $n_{\rm s} = 0.96$ and $H_0 = 100$ $h^{-1}$km s$^{-1}$ Mpc$^{-1}$, where $h = 0.678$ \citep{Planck+2014}, tracing the evolution of $3840^3$ particles from $z$ = 120 to $z$ = 0 in a box of side length 1 $h^{-1}$ Gpc.

The \textsc{\footnotesize{ROCKSTAR}} halo finder \citep{Behroozi_rockstar} was used to identify DM halos and their substructures. This algorithm is 
phase-space-based, searching for overdensities in 
the particle distribution in both position and velocity space, using a linking length of $b=0.28$, guaranteeing that virial spherical overdensities can be determined for even the most ellipsoidal halos. In this context, any overdensities should comprise at least 20 DM particles. According to the selection criteria, central galaxies are identified as galaxies residing in the center of the potential well of host halos. These main structures can host multiple substructures called subhalos, where satellite galaxies reside.

The \sag\ datasets can be found in the \textsc{\footnotesize{CosmoSim}} database\footnote{\url{https://www.cosmosim.org/}}. From the public catalog, we use the positions, star formation rate (parameter `sfr'), the stellar mass as the sum of the mass of stars in the spheroid/bulge and the disk (parameters `mstarspheroid' and `mstardisk', respectively), and the stellar age with the parameter `meanagestars', which is the mean age of the stellar population. For the host halo mass ($\mh$), we use the parameter `halomass', which is the DM halo mass within a radius that contains a mean overdensity of 200 times the critical density of the Universe.

We focus on central galaxies at $z = 0$ hosted by low-mass halos of 10$^{11.6}$ $\leq$ \mh/$\mhalo$ $<$ 10$^{11.8}$. The idea behind this selection is that all the analysis in the paper is at a fixed halo mass 
where assembly bias is strong \citep[e.g.,][]{Gao+2005,LiY+2008,LacernaPadilla2011, WechslerTinker2018}.
The following section shows that these central galaxies show a remarkable signal of galactic conformity at large scales.

%%%%%%%%%%%%%%%%%%%%%%%%%%%%%%%%%
\section{Two-halo galactic conformity}
\label{S_GC}
%%%%%%%%%%%%%%%%%%%%%%%%%%%%%%%%%

We measure the mean quenched fraction $f_{\rm Q}$
of neighboring (secondary) galaxies around central (primary) galaxies at fixed halo mass to assess the galactic conformity at $z=0$. 
We use neighbor galaxies (either centrals or satellites) with stellar mass above $10^{9}~\mstar$. This lower stellar mass limit has been used in other works with the \sag\ catalog
to avoid resolution effects  (e.g., \citealt{Lacerna+2022, Hough+2023, Palma_Lacerna+2025}).
We will refer to the fiducial primary sample with all the central galaxies at fixed halo mass as ``PrimAll''. 
We remove central galaxies in the vicinity of 
massive systems of \mh\ $\geq$ 10$^{13}$ $\mhalo$ out to 5 $\mpc$ to define a primary sample away from massive halos. These are the same samples used by \citet{Lacerna+2022}, who chose this particular scale as a simple representation of the large-scale environment beyond the virial radius of host halos \citep[e.g.,][]{Kuutma+2017, AF+2018}.  
We will refer to this additional sample 
that does not include central galaxies around massive systems as \mbox{``PrimB''}.
Table \ref{tab:parent_samples} lists the respective samples and criteria.
We verified that the stellar mass distribution is very similar between the two parent samples and that the stellar mass-halo mass relation is
relatively flat with halo mass, with an increase in stellar mass of 0.23 dex, which is expected given the narrow halo mass range of 0.2 dex.

%%% Table parent samples
\begin{table}
    \caption{Description and number of galaxies of the parent samples from \sag.}
    \centering
    \renewcommand{\arraystretch}{1.5}
    \begin{tabular}{l p{4.4cm} l}
    \hline\hline
      Parent sample & Criteria & N$_{\rm gal}$ \\
      \hline
      PrimAll   & Central galaxies at $z=0$ with stellar mass above $10^{9}~\mstar$ and hosted by DM halos of 10$^{11.6}$ $\leq$ \mh/$\mhalo$ $<$ 10$^{11.8}$.     & 2 736 547 \\
      PrimB     &  Same as PrimAll, but without central galaxies out to cluster-centric distances of 5 $\mpc$ from massive halos of \mh\ $\geq$ 10$^{13}$ $\mhalo$.  &  1 697 808\\
    \hline  
    \end{tabular}
    \label{tab:parent_samples}
\end{table}

\begin{figure}
\includegraphics[width=\columnwidth]{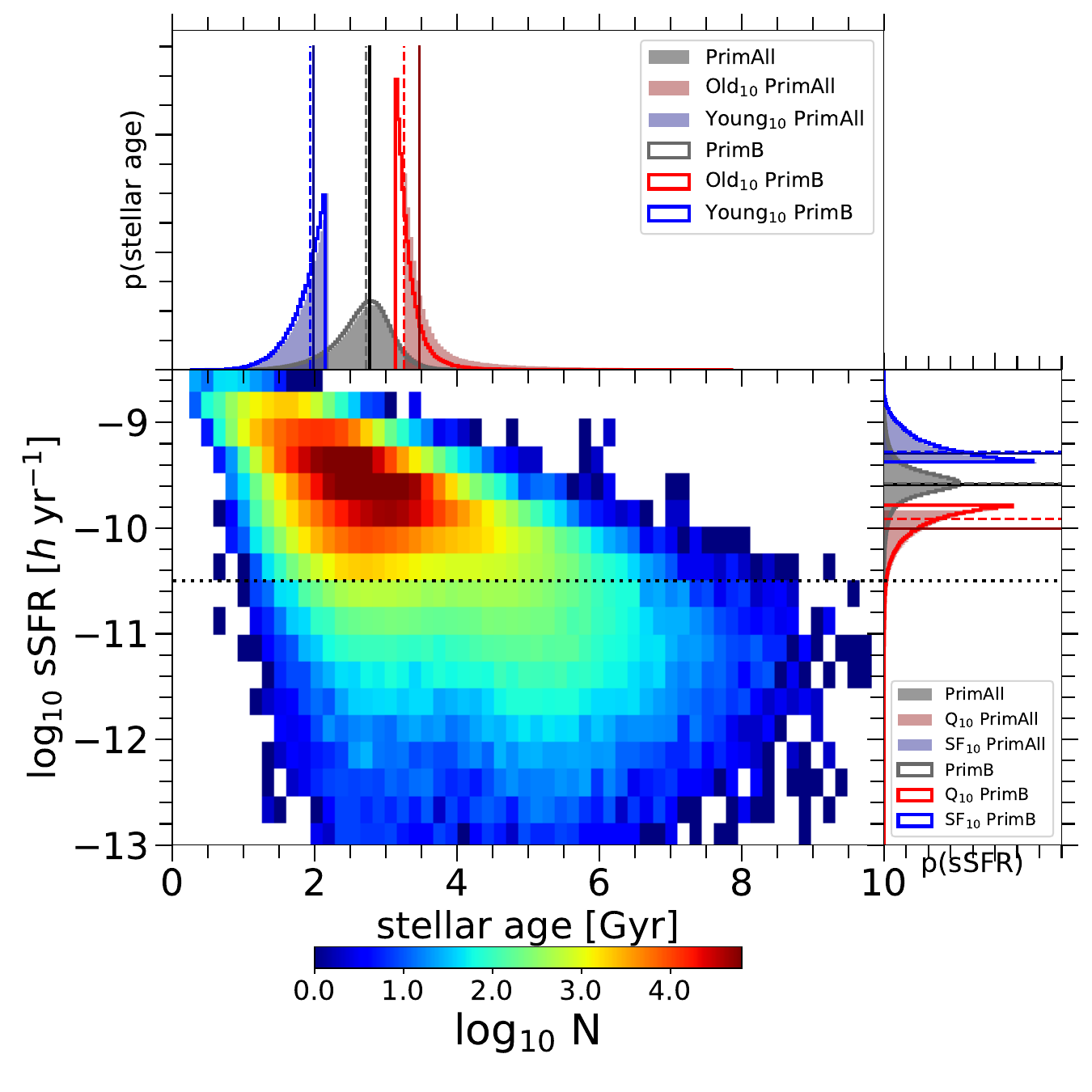}
\caption{
Distribution of sSFR as a function of stellar age for the low-mass central galaxies, represented by the number density of galaxies as indicated in the color bar (main panel). The horizontal dotted line shows the condition used to separate star-forming and quenched galaxies.
The top and right panels show the normalized density distributions of stellar age and sSFR, respectively, for the parent samples and sub-samples as indicated in the legends and described in Tables \ref{tab:parent_samples} and \ref{tab:sub_samples}. The integral of each histogram sums to unity. The lines correspond to the median value for the PrimAll (solid) and PrimB (dashed) samples.
}
\label{fig_ssfr_StellarAge}
\end{figure}

We separated the galaxies into quenched (Q) and star-forming (SF) galaxies at $z = 0$. We consider the
same sSFR cut used in \cite{Lacerna+2022} and \cite{Palma_Lacerna+2025} to split the samples. A galaxy is considered quenched if  {\rm sSFR} $\leq$ 10$^{-10.5}$ $h$ yr$^{-1}$. Otherwise, the galaxy is deemed to be star-forming.
This cut\footnote{Notice that it corresponds to {\rm sSFR} $\sim$10$^{-10.7}$ yr$^{-1}$ with $h = 0.678$, 
close to other cuts commonly used in
the literature, e.g., 10$^{-11}$ yr$^{-1}$ \citep{Wetzel+2012}.}
is chosen because it reproduces well
the bimodality of galaxies in the \sag\ model, as shown in previous studies such as \cite{Brown+2017} and \cite{Cora+2018}. We obtain a fraction of quenched galaxies of about 21 percent in the full catalog. This selection criterion is also applied to the low-mass central galaxies. Table \ref{tab:sub_samples} lists the information of the sub-samples of Q and SF central (primary) galaxies for the cases ``PrimAll'' and \mbox{``PrimB''}.
Fig. \ref{fig_ssfr_StellarAge} shows the distribution of sSFR as a function of stellar age for the low-mass central galaxies. The right-hand panel shows that the parent samples ``PrimAll''
(solid black histogram) and ``PrimB'' (open gray histogram) exhibit very similar sSFR distributions. Their median values of log$_{\rm 10}$(sSFR/$h$ yr$^{-1}$) are -9.59 and -9.57, respectively.

%%% Table sub-samples
\begin{table*}
    \caption{Description and number of sub-samples of central (primary) galaxies for each parent sample.}
    \centering
    \renewcommand{\arraystretch}{1.5}
    \begin{tabular}{cccc}
    \hline\hline
      Sub-sample & Criteria &  PrimAll & PrimB \\
      \hline
      Q   & {\rm sSFR} $\leq$ 10$^{-10.5}$ $h$ yr$^{-1}$  & 40 420 (1.5\%) & 8448 (0.5\%) \\
      SF  & {\rm sSFR} > 10$^{-10.5}$ $h$ yr$^{-1}$  & 2 696 127 (98.5\%) &  1 689 360 (99.5\%)\\
  Q$_{10}$ & at or below 10th percentile of the sSFR distribution & 273 655 (10\%) & 169 781 (10\%)\\
      SF$_{10}$ & at or above 90th percentile of the sSFR distribution & 273 655 (10\%) & 169 781 (10\%) \\
      Young$_{10}$ & at or below 10th percentile of the stellar age distribution & 273 650 (10\%) & 169 780 (10\%)\\
      Old$_{10}$ & at or above 90th percentile of the stellar age distribution & 273 655 (10\%) & 169 781 (10\%) \\
    \hline  
    \end{tabular}
    \label{tab:sub_samples}
\end{table*}

%%%fig sSFR vs stellar age

\begin{figure}
\includegraphics[width=\columnwidth]{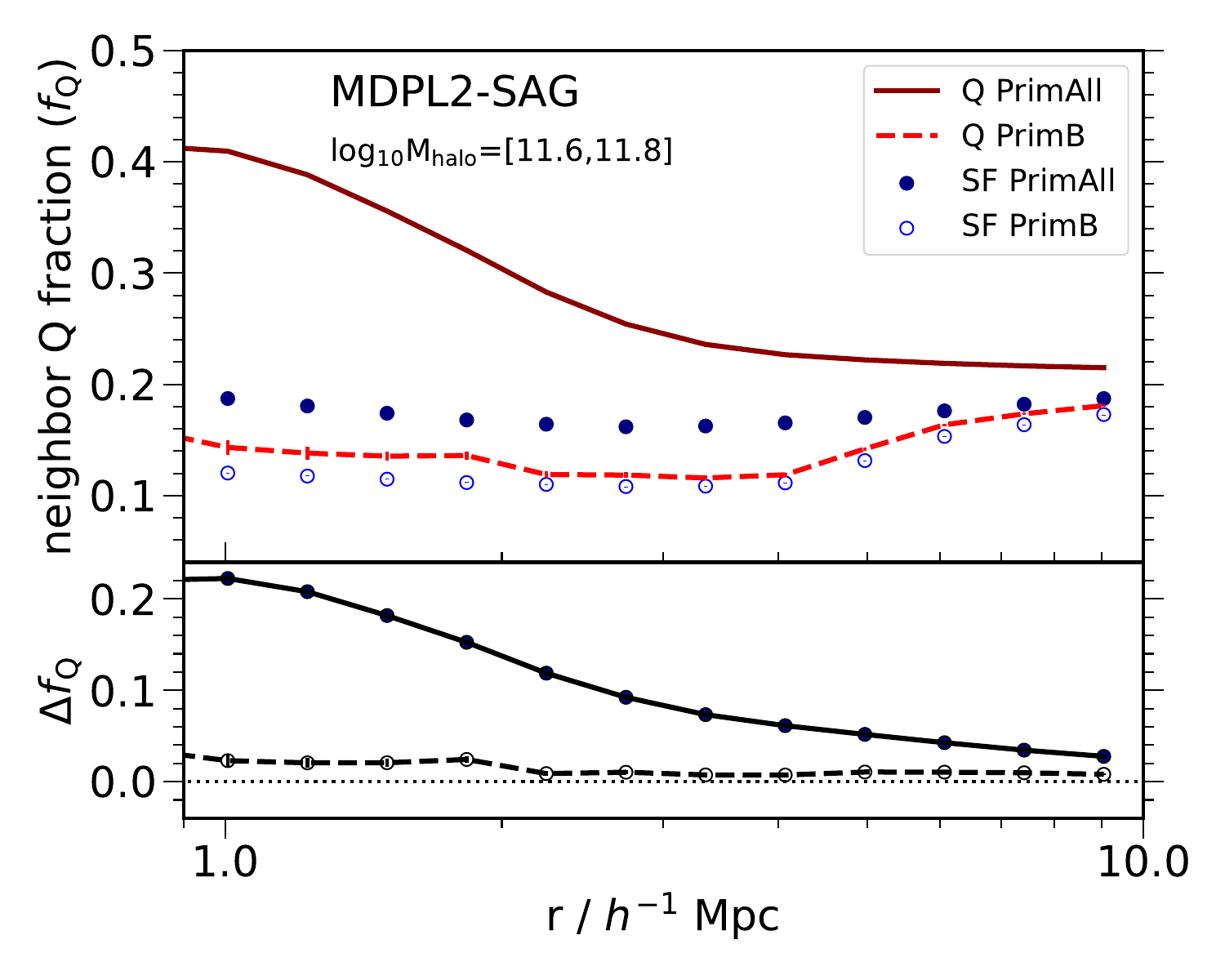}
\caption{
Mean quenched fractions of neighboring galaxies ($f_{\rm Q}$) as functions of the real-space distance from primary galaxies hosted by low-mass halos of 10$^{11.6}$ $\leq$ \mh/$\mhalo$ $<$ 10$^{11.8}$ (main panel).
Primary samples are separated by their sSFR.
The $f_{\rm Q}$ is shown around quenched and star-forming ``PrimAll'' galaxies as a dark red solid line and navy blue solid circles, respectively.
The red dashed line and blue open circles correspond, respectively, to the mean fractions after removing the quenched and star-forming
primary galaxies in the vicinity of halos more massive than 10$^{13}$ $\mhalo$, 
case \mbox{``PrimB''}.
The lower sub-panel shows the difference  
in the mean quenched fractions of neighboring galaxies around quenched and star-forming primary galaxies at fixed halo mass.
The solid line shows the case \mbox{``PrimAll''}, whereas the dashed line is the result obtained for \mbox{``PrimB''}.
The dotted line denotes the case of zero difference,
i.e., no conformity.
}
\label{fig_GC_ssfr_fixedMhalo}
\end{figure}

Figure \ref{fig_GC_ssfr_fixedMhalo} shows the mean quenched fraction of neighbor galaxies around central galaxies hosted by dark matter halos with masses between 10$^{11.6}$  and 10$^{11.8}$ $\mhalo$. This figure is similar to that presented in \cite{Lacerna+2022} and is also shown in this work for self-consistency.
The dark red solid line and navy blue solid circles correspond to $f_{\rm Q}$ around 
quenched and star-forming central galaxies in the primary sample 
\mbox{``PrimAll’'}, respectively.
The red dashed line and blue open circles show, respectively, the result after removing the quenched and star-forming central galaxies in the vicinity of halos more massive than
10$^{13}$ $\mhalo$ from the primary sample, case 
``PrimB’'.
The errors in the estimation of the mean fractions are calculated using $120$ jackknifes
\citep[e.g.,][]{Zehavi+2002, Norberg+2009}.
Error bars in the mean fractions are estimated by using the diagonal of the covariance matrix. 
Given the large number of galaxies, the error bars are small enough to be imperceptible in the figure.

The lower sub-panel of Fig. \ref{fig_GC_ssfr_fixedMhalo} shows the two-halo conformity signal for the two cases, measured as the difference of the mean quenched fractions of neighboring galaxies around quenched and star-forming primary galaxies ($\Delta f_{\rm Q}$) at fixed halo mass. 
The signal is strong in the fiducial case (solid line) where the fraction of quenched neighbor galaxies at distances of $\sim 1~\mpc$ from primary galaxies is much higher around quenched centrals than around star-forming centrals, with $\Delta f_{\rm Q}$ = 0.22. This correlation decreases with distance from the central galaxies, becoming quite low at distances $r \gtrsim 6~\mpc$ with $\Delta f_{\rm Q} \lesssim 0.04$. In contrast, the two-halo conformity signal of the case ``PrimB’' (dashed line)
is always very low at distances $r \gtrsim 1~\mpc$ with $\Delta f_{\rm Q} \lesssim 0.02$. As shown by \cite{Lacerna+2022}, this result is not an artifact of removing galaxies from the primary sample. The low-mass central galaxies in the vicinity of galaxy groups and clusters are responsible for the majority of the conformity signal at megaparsec scales and show a larger neighbor quenched fraction than other centrals since, when removing these, the $f_{\rm Q}$ is also lower. 
In the next section, we will study their potential contribution to the assembly-type bias. 

%%%fig conformity sSFR at fixed Mhalo

%%%%%%%%%%%%%%%%%%%%%%%%%%%%%%%%%
\section{Assembly-type bias}
\label{S_AB}
%%%%%%%%%%%%%%%%%%%%%%%%%%%%%%%%%

In this section, we explore the contribution of low-mass central galaxies around massive groups and clusters to the assembly-type bias signal. As shown in the previous section, central galaxies hosted by low-mass halos near massive halos can greatly explain most of the two-halo conformity signal. Similarly to that approach, we first measure the two-point correlation functions for the fiducial case ``PrimAll'', and then for the case ``PrimB''
that excludes the low-mass central galaxies in the vicinity of massive structures.

We use the software \textsc{CORRFUNC} \citep{corrfunc, 10.1007/978-981-13-7729-7_1, SinhaLehman2020} to measure the two-point correlation functions using the Landy-Szalay estimator \citep{LS}. 
Given the geometrical effect of removing galaxies out to cluster-centric distances of 5 $\mpc$, we have constructed a random catalog for the case ``PrimB''
that considers this effect. First, we created a fiducial random catalog within the simulation box that is three times the number of primary galaxies, producing a catalog of $\sim$8,200,000 random objects. Then, we use the positions of groups and clusters in the simulation to remove the random points at cluster-centric distances $\le 5~\mpc$ in real space. The random catalog of the case ``PrimB'' is made of the remaining random points (about $\sim7,000,000$ objects). 
The error bars in the clustering estimations of this section are calculated using the diagonal of the covariance matrix from the bootstrap method \citep[e.g.,][]{Norberg+2009} by randomly resampling each galaxy sample 25 times with replacement.

%%%%
\subsection{Quenched and star-forming samples for a defined cut in {\rm sSFR}}
\label{subS_AB_sSFR}
%%%%

The aim of this sub-section is to assess the contribution of the galaxies that produce the two-halo conformity to the assembly-type bias signal using the same conditions of Section \ref{S_GC}, 
i.e., we estimate the two-point correlation functions for the samples of quenched and star-forming primary galaxies defined in the previous section using a cut in {\rm sSFR} of 10$^{-10.5}$ $h$ yr$^{-1}$. 
Given the low-mass regime of \mh\ between 10$^{11.6}$  and 10$^{11.8}$ $\mhalo$, the fraction of SF central galaxies is much higher than the fraction of Q central galaxies with the same host halo mass (98.5\% and 1.5\%, respectively, see Table \ref{tab:sub_samples}).
Sect. \ref{subS_AB_sSFR_10perc} will show the results for samples with the same number of quenched and star-forming central galaxies.

\subsubsection{Fiducial case ``PrimAll''}
\label{subS_AB_PrimAll}
%%

%%%fig assembly-type bias with sSFR at fixed Mhalo
\begin{figure}
\includegraphics[width=\columnwidth]
{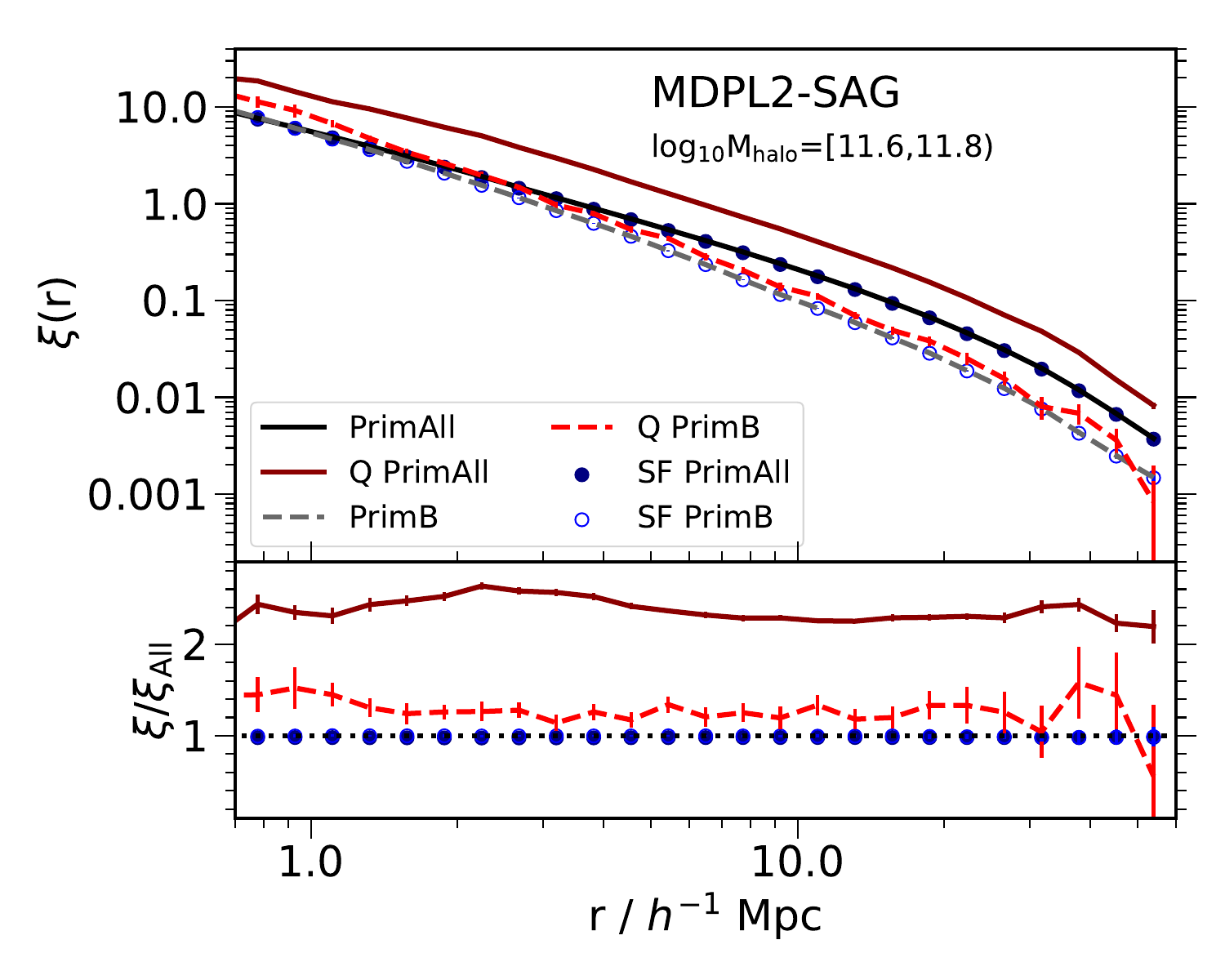}
\caption{
The two-point correlation functions of primary galaxies hosted by dark matter halos with masses of 10$^{11.6}$ $\leq$ \mh/$\mhalo$ $<$ 10$^{11.8}$.
The autocorrelation function for the case with all the central galaxies in the primary sample, ``PrimAll'', is shown as a solid black line, whereas the dark red solid line and 
navy blue circles correspond to the cross-correlation functions of quenched and star-forming primary galaxies, respectively.
On the other hand, the dashed gray line corresponds to the autocorrelation function after removing the central galaxies in the vicinity of halos more massive than 10$^{13}$ $\mhalo$ from the primary sample, case ``PrimB''.
The red dashed line and open blue circles correspond to the cross-correlation functions of quenched and star-forming primary galaxies in the case \mbox{``PrimB''}, respectively. The dashed lines and open circles are then the results after removing the galaxies that mostly contribute to the two-halo galactic conformity.
The sub-panel shows the ratio between the correlation function of 
each sub-sample
and the respective parent sample. 
The solid and dashed lines are
for the cases Q ``PrimAll'' and Q ``PrimB'', respectively. The solid and open circles are for the cases SF ``PrimAll'' and SF ``PrimB'', respectively. There is an overlap in the circles because both ratios are close to unity.
}
\label{fig_AB_ssfr_fixedMhalo}
\end{figure}

Figure \ref{fig_AB_ssfr_fixedMhalo} shows the two-point cross-correlation functions between the quenched primary galaxies and the full sample of primary galaxies (dark red solid line) and between star-forming primary galaxies and the full sample of primary galaxies (navy blue solid circles) in the case ``PrimAll''. They correspond to the fiducial case of central galaxies hosted by low-mass dark matter halos between $M_{\rm h}$ = 10$^{11.6}$ and 10$^{11.8}$ $\mhalo$. The overall clustering of central galaxies is expected to be dominated by star-forming galaxies because they correspond to the majority of the centrals in the halo mass range explored in this paper. This behavior is demonstrated when we show the autocorrelation function of all the primary galaxies in the fiducial case as a black solid line, with a clustering almost the same as the cross-correlation of the SF primaries (navy blue circles). On the other hand, the clustering of quenched central galaxies is higher than that of star-forming centrals hosted by halos of the same mass. This result is also seen in the solid lines of the bottom panel, which shows the ratio between the correlation function of quenched
centrals and the total population (dark red solid line) and between the star-forming centrals and the total population (navy blue solid circles) of the fiducial case. The ratio is a factor above 2 for the quenched central galaxies out to 50 $\mpc$, which implies a strong secondary bias on this population because the galaxy clustering depends on the sSFR at fixed halo mass.

%%%%
\subsubsection{Case ``PrimB''}
%%%%

In Sect. \ref{S_GC}, we mentioned the significant contribution of central galaxies in the vicinity of massive halos in the two-halo galactic conformity signal. To assess the contribution of these galaxies in the fiducial signal of the assembly-type bias shown in Sec.
\ref{subS_AB_PrimAll}, we repeat the process of removing those galaxies from the primary sample, i.e., we estimate the two-point correlation functions for the case ``PrimB''.
The cross-correlation function between the quenched (star-forming) primary galaxies and the primary galaxies for this case is shown in the red dashed line (blue open circles) in the main panel of Fig. \ref{fig_AB_ssfr_fixedMhalo}. The autocorrelation function of the primary galaxies of the case ``PrimB'' is shown as a gray dashed line. Star-forming galaxies still dominate the overall clustering of central galaxies because they correspond to 99.5\% of the centrals in this case (see Table \ref{tab:sub_samples}).

A difference compared with the fiducial case is that the clustering decreases for the quenched and star-forming populations, even when the halo mass range is the same, probably due to the removal of galaxies in the vicinities of massive halos. Furthermore, the difference in the clustering between the quenched and star-forming galaxies of the same halo mass is remarkably reduced, which is also shown by the correlation function ratios in the bottom panel of the figure 
in red dashed line and blue open circles. The average ratio in the two-halo regime between 1 and 50 $\mpc$ is 1.28
for the quenched central galaxies, which is a decrease of 
$\sim$5 times in the relative assembly-type bias on this population when low-mass central galaxies near massive systems are not considered in the clustering estimations.

%%%%
\subsection{Equal samples of quenched and star-forming central galaxies}
\label{subS_AB_sSFR_10perc}
%%%%

The results presented in Sect. \ref{subS_AB_sSFR} are strongly weighted by the dominant number of central galaxies in the star-forming samples compared to the quenched samples. One may argue that comparing the cross-correlations between both samples is not ``fair'' due to the differences in number.  
Studies of assembly bias typically use samples of similar or equal numbers of halos/galaxies in the extremes of the distributions of secondary parameters. To evaluate the consistency of the results of Sect. \ref{subS_AB_sSFR} 
with the literature, we here use the 10 percent of the most quenched central galaxies (Q$_{\rm 10}$) and the 10 percent of the most star-forming central galaxies (SF$_{\rm 10}$) in the halo mass range of  10$^{11.6}$ and  10$^{11.8}$ $\mhalo$.
This way, we have the same number of central galaxies in both samples, including and excluding galaxies close to massive halos. There are 273,655 quenched central galaxies and the same number of star-forming centrals in the case ``PrimAll'', while this number decreases to 169,781
central galaxies for each sample of quenched and star-forming galaxies in the case ``PrimB'' (see Table \ref{tab:sub_samples}).
As well as the case of the parent samples, the fraction of central galaxies with sSFR $\leq$ 10$^{-10.5}$ is higher for the Q$_{\rm 10}$ ``PrimAll'' sub-sample than the Q$_{\rm 10}$ ``PrimB'', with quenched fractions of 0.15 and 0.05, respectively. This result is in agreement with \citet{Palma_Lacerna+2025}, who find a larger number of quenched central galaxies near massive halos than in the field at fixed stellar mass because ``PrimB'' samples correspond to "field" central galaxies, whereas "PrimAll" samples include central galaxies near massive groups and clusters.

\begin{figure*}
\includegraphics[width=0.5\textwidth]
{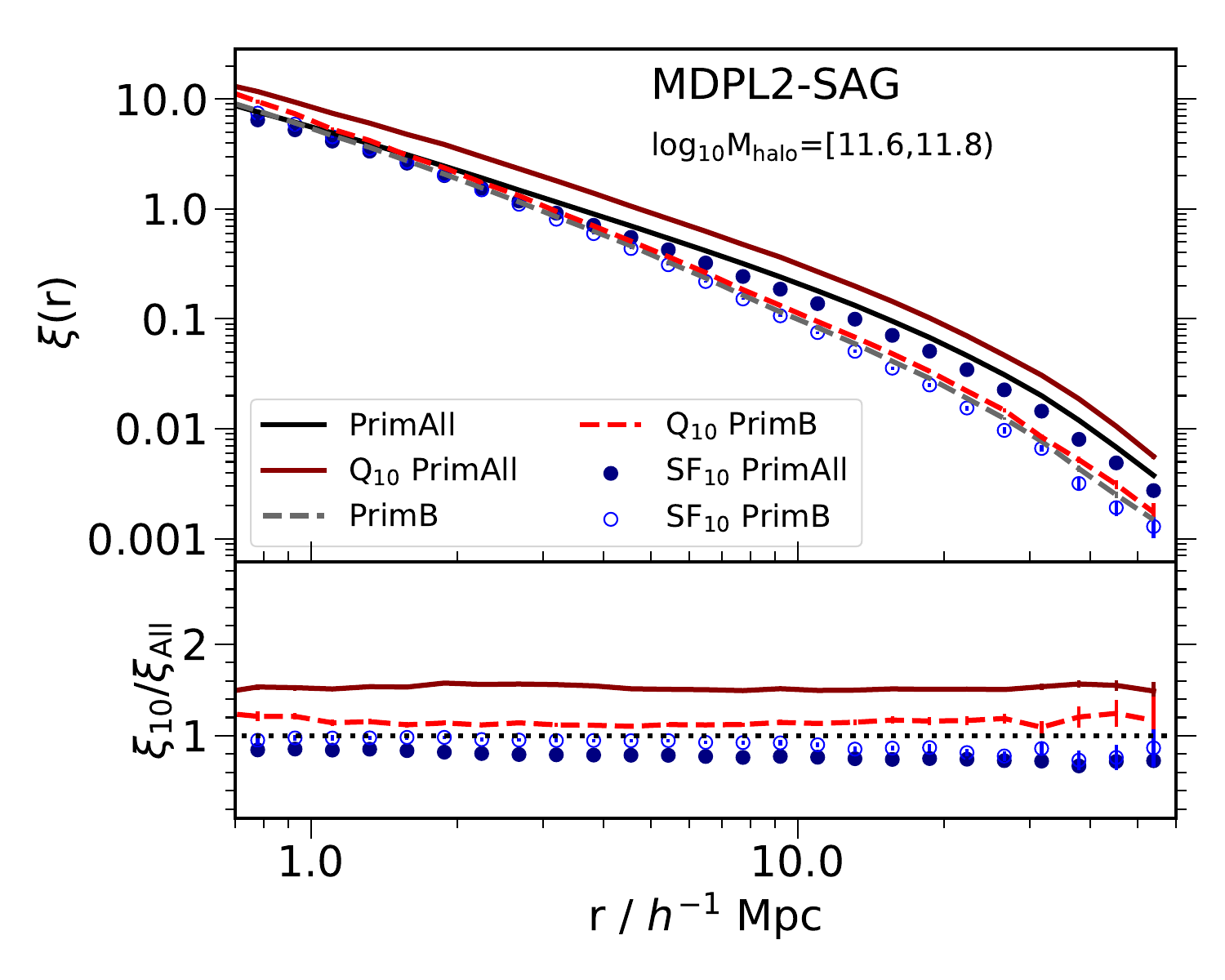}
\includegraphics[width=0.5\textwidth]
{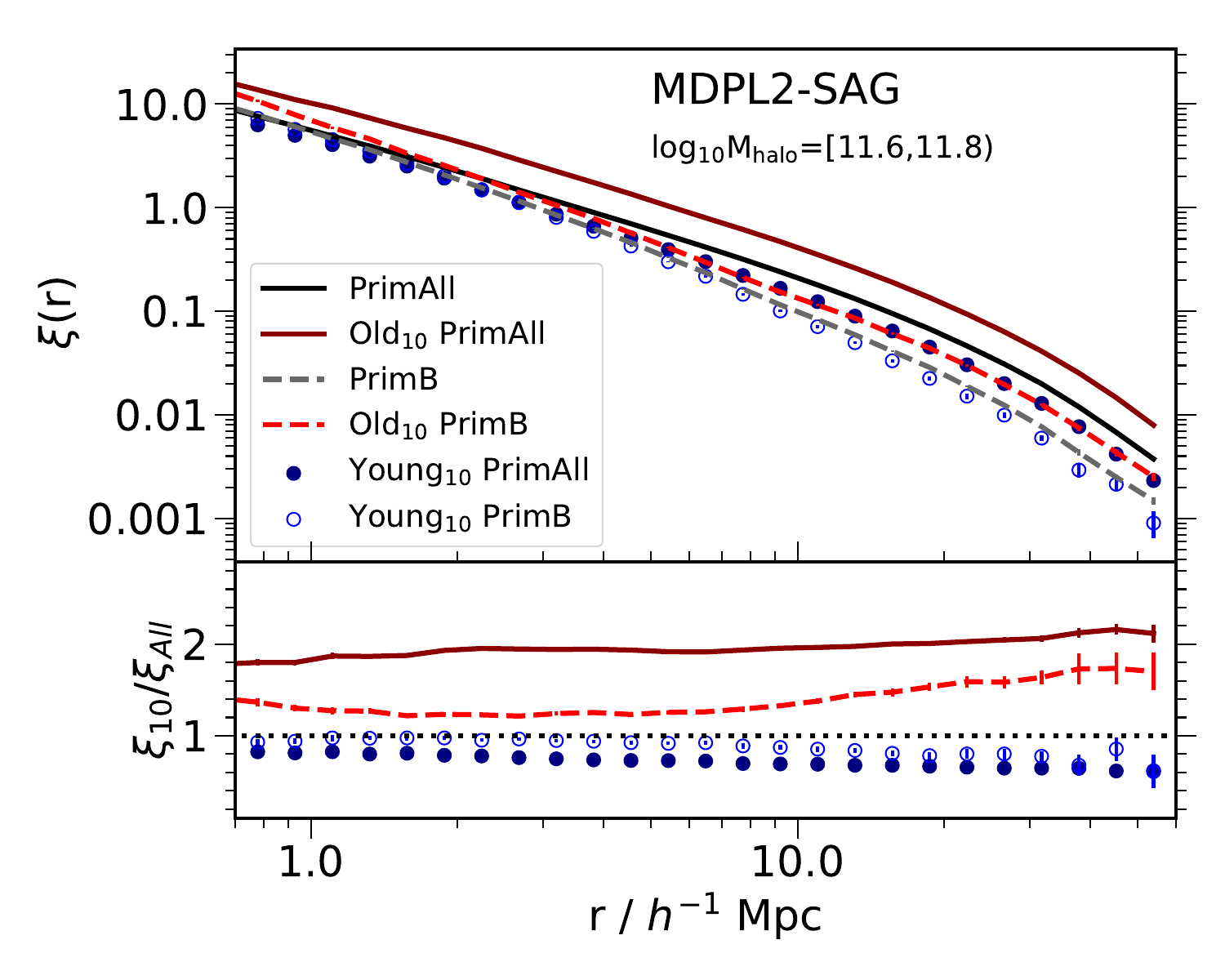}
\hspace*{0.4cm}\includegraphics[width=0.49\textwidth]
{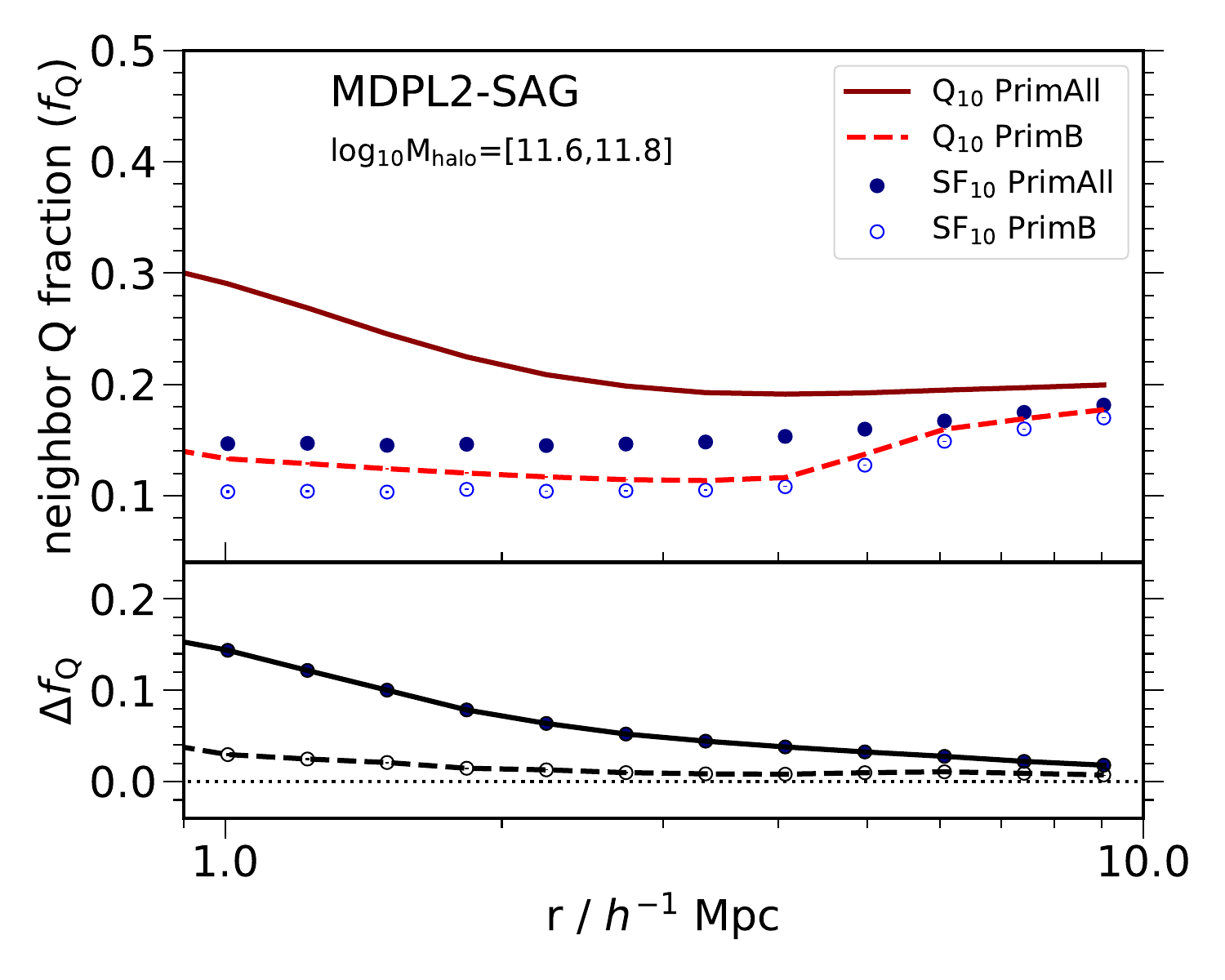}
\hspace*{0.15cm}\includegraphics[width=0.49\textwidth]
{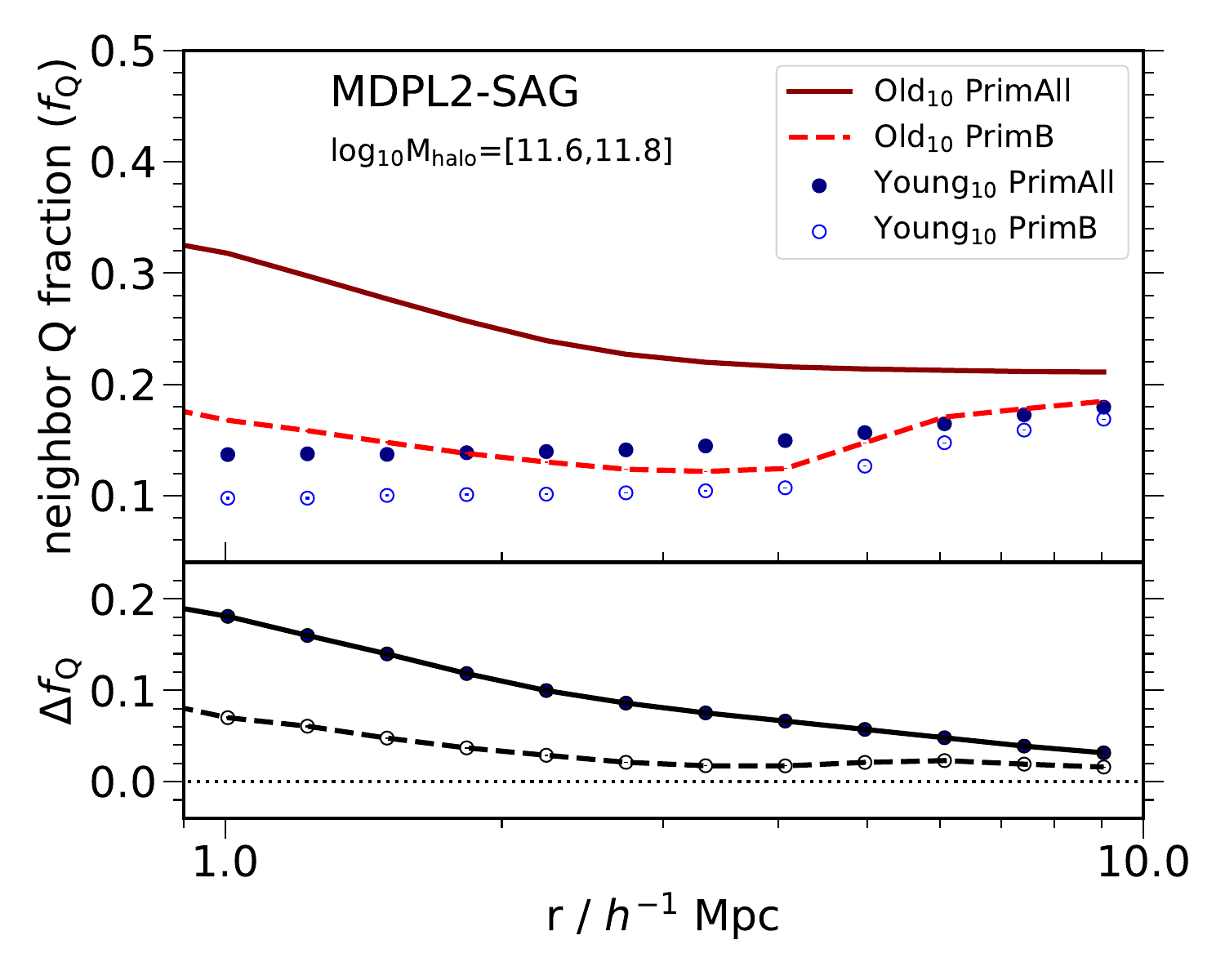}
\caption{Top left (right): Same as Fig. \ref{fig_AB_ssfr_fixedMhalo}, but the cross-correlations 
are between 
the most quenched (oldest) or the 10\% of the most star-forming (youngest) central galaxies and the respective 
parent sample of primary galaxies.
Bottom: Same as Fig. \ref{fig_GC_ssfr_fixedMhalo}, but the primary samples are separated between the 10 percent of the most quenched and the 10 percent of the most star-forming central galaxies (left) and between the  10 percent of the oldest and 10 percent of the youngest central galaxies (right).
}
\label{fig_AB_GC_10perc}
\end{figure*}

The top-left panel of Fig.
\ref{fig_AB_GC_10perc} shows the cross-correlation functions of these sub-samples for the cases ``PrimAll'' and \mbox{``PrimB''}. The autocorrelations of the parent samples
are also shown, which are the same as Fig. \ref{fig_AB_ssfr_fixedMhalo} by definition, and we use the same symbols and line styles as in that figure. It is now clear the separation in the clustering of both the most quenched and the most star-forming central galaxies with respect to the autocorrelation function of all the central galaxies at fixed halo mass in the case ``PrimAll'', i.e., an assembly-type bias, with a higher and lower clustering respectively, as also shown in the solid %lines
line and solid circles in the sub-panel of the figure.
This behavior is due to the high star formation sample including 
fewer SF galaxies than in the previous case, 
but also with slightly higher sSFRs on average,
and the least star-forming counterpart containing the quiescent galaxies of Section 4.1. 
The clustering of the $10\%$ most quenched galaxies is, on average, about 50\% higher than all the central galaxies between 1 and 50 $\mpc$, whereas it is about 25\% lower for the most star-forming central galaxies.

When central galaxies in the vicinity of galaxy groups and clusters are not used in the estimations, the clustering of the most quenched galaxies is about 15\% higher, whereas it is 10\% lower for the most star-forming galaxies, on average between 1 and 50 $\mpc$ with respect to all the central galaxies in the case \mbox{``PrimB''}.
The above means a decrease of a factor of three
in the relative assembly-type bias for equal samples of quenched and star-forming central galaxies. We notice a slight increase in the ratios of ``PrimB'' galaxies at scales above $\sim$ 10 $h^{-1}$ Mpc.
We will come back to this further below.   

We confirmed that the distributions for the respective quenched and star-forming sub-samples are similar between the cases ``PrimAll'' and ``PrimB'' (see the right-hand panel of Fig. \ref{fig_ssfr_StellarAge}). The median log$_{10}$(sSFR / $h$ yr$^{-1}$) is -10.00 (-9.29) and -9.91 (-9.27) for the 10\% of the most Q (SF) centrals in the ``PrimAll'' and ``PrimB'' samples, respectively. Therefore, the differences in the clustering between the “PrimAll” and “PrimB” are not due to different sSFR of the chosen quenched and star-forming samples.

The assembly-type bias is strongly reduced in the case \mbox{``PrimB''} when separating the samples according to their sSFR. We observe a similar decrease 
in the two-halo conformity signal using these sub-samples of primary galaxies.
The bottom-left panel of Fig. \ref{fig_AB_GC_10perc}
shows the mean quenched fractions of neighboring galaxies around primary galaxies separated for the 10 percent of the most quenched (Q$_{\rm 10}$) and the 10 percent of the most star-forming (SF$_{\rm 10}$) central galaxies in the same host halo mass range. In this way, we can assess whether the conformity signal varies in the same manner as the assembly-type bias signal 
under these conditions. We maintain the definition that secondary galaxies are quenched if {\rm sSFR} $\leq$ 10$^{-10.5}$ $h$ yr$^{-1}$.  
The quenched fractions of neighboring (secondary) galaxies decrease around primary galaxies in the case ``PrimAll'' (dark red solid line and navy blue solid circles) compared to the results in Fig. \ref{fig_GC_ssfr_fixedMhalo}. The conformity signal (black solid line) also decreases from $\Delta f_{\rm Q}$ = 0.22 to 0.15 at separations of $r \sim 1~\mpc$, but it is relatively strong out to distances $r \lesssim 4~\mpc$ with $\Delta f_{\rm Q} > 0.04$. This lower signal is expected because 
10 percent of the most quenched primary galaxies  
include centrals with {\rm sSFR} > 10$^{-10.5}$ $h$ yr$^{-1}$. On the other hand, the conformity signal for the case ``PrimB'' (black dashed line) is similar to that shown in Fig. \ref{fig_GC_ssfr_fixedMhalo}, although with slightly higher values of $\Delta f_{\rm Q} \sim 0.03$ at separations between 1 and 1.5 $\mpc$.

When studying conformity using the exact percentiles in sSFR, the conformity signal for the fiducial case is also present, but diminished with respect to using quenched and star-forming samples.
It seems that low-mass central galaxies in the vicinity of groups and clusters, responsible for the ``standard'' two-halo conformity, might be able to
explain the assembly-type bias partially, but not all of it. The following section studies the signal strength based on another galaxy property, the stellar age.

%%% figs assembly-type bias AND conformity using 10% sSFR and stellar age

%%%%
\subsection{Separating samples of central galaxies according to their stellar ages}
\label{subS_AB_Sages_10perc}
%%%%

We have shown that central galaxies responsible for the two-halo conformity in the low-mass regime can mostly account for the assembly-type bias signal at separations of dozens of Mpc. The signals measured in Fig. \ref{fig_AB_ssfr_fixedMhalo} and 
Fig. \ref{fig_AB_GC_10perc} (top-left panels)
were estimated after separating the sample of central galaxies at fixed halo mass according to their sSFR. However, the sSFR might not be the best tracer of the host halo formation time, a secondary halo property that exhibits strong assembly bias in low-mass DM halos, since it quantifies the level of recent star formation in galaxies. 

As suggested by \citet{LacernaPadilla2011}, a better observational tracer for the halo formation time at fixed halo mass is the stellar age. Therefore, in this section, we measure the signal of assembly-type bias for the fiducial case ``PrimAll'' and the case without central galaxies in the vicinity of massive structures \mbox{``PrimB''}
using the stellar age of the central galaxies. For each case, we separate the 10 percent of the oldest and the 10 percent of the youngest galaxies (see Table \ref{tab:sub_samples}).
The correlation functions are shown in 
the top-right panels of Fig. \ref{fig_AB_GC_10perc}.
The difference in clustering between the oldest and youngest galaxies in the fiducial case (dark red solid line and navy blue solid circles) is much stronger than between the most quenched and star-forming galaxies
(top-left panels of Fig. \ref{fig_AB_GC_10perc}),
which confirms that the stellar age is a better proxy for the halo formation time than the sSFR.
The clustering of the oldest galaxies is, on average, about 2 times higher than for the case of all the
central galaxies between 1 and 50 $\mpc$, whereas it is about
30\% lower for the youngest central galaxies. In the case \mbox{``PrimB''} (dashed lines and open circles), the clustering between the oldest and youngest galaxies drastically reduces between 1 and $\sim$10 $\mpc$ by a factor of $\sim$4. However, the assembly-type bias becomes stronger at larger separations in a clearer way than we found for samples separated according to their sSFR.  We will concentrate on a possible explanation for this in the following subsection.

We confirmed that the two-halo conformity signal using these sub-samples of primary galaxies also correlates with the assembly-type bias of
the top-right panels of Fig. \ref{fig_AB_GC_10perc}.
The bottom-right panel of Fig. \ref{fig_AB_GC_10perc}
shows the mean quenched fractions of neighboring galaxies around primary galaxies separated between the 10 percent oldest (Old$_{\rm 10}$) and the 10 youngest (Young$_{\rm 10}$) central galaxies in the same host halo mass range. The quenched fractions of neighboring galaxies also decrease with the distance from primary galaxies in the case ``PrimAll'' compared to the results in Fig. \ref{fig_GC_ssfr_fixedMhalo}. The conformity signal (black solid line) decreases from $\Delta f_{\rm Q}$ = 0.22 to 0.18 at separations of $r \sim 1~\mpc$. It is relatively strong at distances $r \lesssim 7~\mpc$ with $\Delta f_{\rm Q} \gtrsim 0.04$. Therefore, the conformity signal is stronger after separating the fiducial primary sample by stellar age than by sSFR.
This result is consistent with the assembly-type bias being stronger using the stellar age compared to the sSFR.

The conformity signal is notably reduced for the samples \mbox{``PrimB''}, but the residual signal is higher when separating for the stellar ages than for the sSFR.
In contrast to the bottom-left panel of Fig. \ref{fig_AB_GC_10perc},
the bottom-right panel shows an evident two-halo conformity signal for the case ``PrimB'' at separations $r \lesssim 2~\mpc$. This result is mainly produced by the higher quenched fraction of neighbors around the oldest primary galaxies (red dashed line). We notice that 10 percent of the oldest ``PrimB'' galaxies have a median stellar age of 3.3 Gyr (see the red dashed line in the top panel of Fig. \ref{fig_ssfr_StellarAge}), which also include many star-forming galaxies  with {\rm sSFR} > 10$^{-10.5}$ $h$ yr$^{-1}$, which are not part of the 10 percent of the most quenched galaxies. Nonetheless, the conformity signal decreases notably in the case ``PrimB'' compared to the fiducial case ``PrimAll'' at $r \geq 1~\mpc$.

We checked that the change in clustering is not due to a substantial change in the stellar age distributions 
of the oldest or youngest primary galaxies between the cases 
``PrimAll'' and \mbox{``PrimB''} (see the top panel of Fig. \ref{fig_ssfr_StellarAge}).
The median stellar age of the 10 percent of the oldest galaxies in ``PrimAll'' is 3.47 Gyr, whereas it is 3.26 Gyr in \mbox{``PrimB''}. In the case of the 10 percent of the youngest central galaxies, the median age is 1.98 Gyr in ``PrimAll'', whereas it is 
1.93 Gyr in \mbox{``PrimB''}.
The stellar ages are slightly younger when the central galaxies in the vicinity of massive structures are not considered compared with the fiducial case, but not more than 200 Myr on average.  This result shows that the different amplitudes of assembly bias between the  ``PrimAll''
and ``PrimB'' are not due to the difference in stellar age of the chosen old and young subsamples.

%%%%%%%%%%%%%%%%%%%%%%%%%%%%%%%%%
\subsection{Effect of the removal of galaxies around massive halos}
\label{Sdiscussion}
%%%%%%%%%%%%%%%%%%%%%%%%%%%%%%%%%

We found that the low-mass central galaxies in the vicinity of massive galaxy groups and clusters have, in addition to their contribution to the two-halo galactic conformity, an important role in the assembly-type bias of central galaxies at fixed halo mass at a separation of dozens of megaparsecs. The signal of assembly-type bias decreases by a factor of three to four if they are not included in the clustering estimations (case ``PrimB''), with a possible slight increase around 10 $\mpc$. In particular, the results of 
the top-right panel of Fig. \ref{fig_AB_GC_10perc},
when separating between the 10\% of the oldest and the 10\% of the youngest galaxies, show the most noticeable increase in the signal at these large separations, reaching a clustering similar to the fiducial case (``PrimAll'') at $r > 50~\mpc$ within the error bars.

\begin{figure}
\includegraphics[width=\columnwidth]{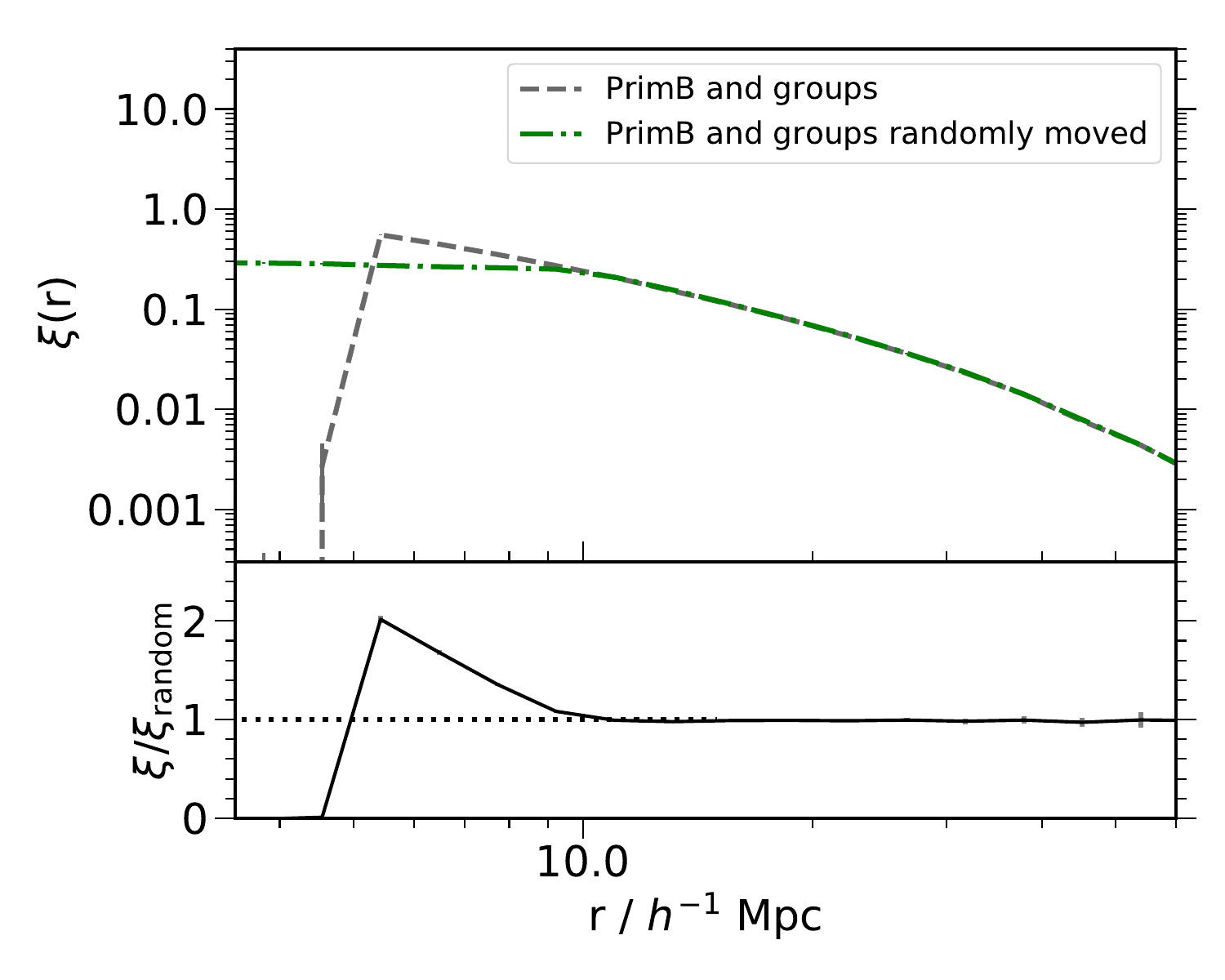}
\caption{Cross-correlation function between the ``PrimB'' sample and the groups and clusters with masses above $10^{13}~\mhalo$ (gray dashed line). The green dot-dashed line corresponds to the cross-correlation that uses random positions of massive halos with a vector of 5 $\mpc$ length from the original positions.
The sub-panel shows the ratio between the correlation functions using the original and random positions of massive halos. They are the same at scales larger than 10 $\mpc$.
}
\label{fig_crossAB_PrimB_clusters}
\end{figure}

The case ``PrimB'' is based on removing central galaxies near massive halos out to cluster-centric distances of 5 $\mpc$.
It is of interest to check whether there is a particular set of scales in which the clustering of the remaining central galaxies is affected by the massive halos located at the center of the removed regions.
To quantify the impact on the selection of this region, we randomly move the position of dark matter halos with masses $\mh \geq 10^{13} \mhalo$ by a fixed distance of 5 $\mpc$ from the original positions.
We then estimate the cross-correlation function between the ``PrimB'' sample and the massive halos with the original and the new halo positions. 
The result is shown in Fig. \ref{fig_crossAB_PrimB_clusters}.
The clustering with the original and random positions of massive halos (gray dashed and green dot-dashed lines, respectively)
differs at separations smaller than 10 $\mpc$, with the clustering of the sample with displaced halos being lower than the original one, until they tend to similar amplitudes at separations larger than 10 $\mpc$. This is the same scale on which we observe an increase in the assembly-type bias strength in the case ``PrimB''.   

The same clustering at separations larger than 10 $\mpc$, regardless of the exact position of massive halos, means that there is a large-scale correlation with them that is still present in the correlation function even when massive halos are moved to a distance of $5~\mpc$ from their original positions.
Therefore, although the short-range effect of assembly-type bias is removed in ``PrimB'', there is still an excess of clustering due to the presence of 
massive halos at larger scales contributing to the signal.  
The increase of the assembly-type bias signal at large separations is less clearly seen in the case where samples are divided according to their sSFR, but in these cases, the error bars are large at large separations.

It is interesting to note that this type of selection could produce a scale-dependent bias. 
It is possible that some galaxy selections targeting, for instance, samples of emission-line galaxies,  could be affected in a similar way.  In particular, emission-line galaxies could avoid  high-density regions dominated by quenched objects but still be affected by modes produced by massive clusters of galaxies.

%%%%%%%%%%%%%%%%%%%%%%%%%%%%%%%%%
\section{Discussion and conclusions}
\label{Sconclusions}
%%%%%%%%%%%%%%%%%%%%%%%%%%%%%%%%%

Environmental effects in the vicinity of massive galaxy groups and clusters seem to cause a significant fraction
of two apparently
different phenomena: two-halo galactic conformity and assembly-type bias. 
The former is a term used to describe the correlation between color or star formation activity in low-mass central galaxies
and neighboring galaxies in adjacent halos at separations of several megaparsecs, whereas the latter is the secondary halo bias reflected in the large-scale clustering of central galaxies.
We evaluated the actual level of equivalence between both phenomena using the \sag\ galaxy catalog constructed by combining the semi-analytic model of galaxy formation {\sc sag} with the  dark matter only MultiDark Planck 2 (\textsc{mdpl2}) cosmological simulation. 

We focused on synthetic central galaxies at $z = 0$ hosted by low-mass halos of 10$^{11.6}$ $\leq$ \mh/$\mhalo$ $<$ 10$^{11.8}$ because it is a mass range where the assembly-type bias has been reported to be strong. We referred to the fiducial primary sample with all the central galaxies at fixed halo mass as ``PrimAll''. We used an additional sample of primary galaxies away from massive halos referred to as ``PrimB'', which does not include central galaxies around massive systems of \mh\ $\geq$ 10$^{13}$ $\mhalo$ out to a cluster-centric distance of 5 $\mpc$. 

The mean fraction of quenched neighboring galaxies at distances between $\sim$1 and 6 $\mpc$ from primary galaxies is much higher around quenched centrals than around star-forming centrals in the fiducial case, which implies a strong signal of conformity at fixed halo mass. In contrast, the two-halo conformity signal of the case ``PrimB’' is always very low at distances $r \gtrsim 1~\mpc$ (Fig. \ref{fig_GC_ssfr_fixedMhalo}). We measured the two-point cross-correlation functions for the fiducial case ``PrimAll'' and for the case ``PrimB”, which excludes low-mass central galaxies in the vicinity of massive structures. The ratio between the correlation function of quenched ``PrimAll'' galaxies and the total population of central galaxies at fixed halo mass is a factor greater than 2,
implying a strong secondary bias on this population. We found that the relative assembly-type bias decreases about 5 times on average between 1 and 50 $\mpc$ when low-mass central galaxies near massive systems are not considered in the clustering estimations (Fig. \ref{fig_AB_ssfr_fixedMhalo}). Therefore, both the galactic conformity and the assembly-type bias are strongly produced by quenched low-mass central galaxies near massive halos.
 
The fraction of quenched central galaxies is 1.5\% in the halo mass regime studied (Table \ref{tab:sub_samples}).
When separating the samples of quenched and star-forming galaxies 
with the same fraction (10\% in our case) as is usually done in the literature on the secondary bias, the relative assembly-type bias decreases about 3 times 
in the case ``PrimB” compared to the fiducial case ``PrimAll'' 
(top-left panel of Fig. \ref{fig_AB_GC_10perc}).
Still, the case \mbox{``PrimB''} shows that the clustering of the most quenched galaxies is about 15\% higher, whereas it is 10\% lower for the most star-forming galaxies, on average between 1 and 50 $\mpc$ with respect to all the central galaxies.
Therefore, low-mass central galaxies in the vicinity of groups and clusters, responsible for the two-halo conformity, might be able to explain the assembly-type bias partially, but not all of it. 
When looking at the conformity signal for these sub-samples, we do find a strong conformity signal for ``PrimAll'' and, after removing galaxies close to massive halos, the conformity signal
between 1 and 1.5 $\mpc$
decreases less than for the case \mbox{``PrimB''} shown in Fig. \ref{fig_GC_ssfr_fixedMhalo}  
We show the corresponding conformity signals in the 
bottom-left panel of Fig.
\ref{fig_AB_GC_10perc}.
The higher two-halo conformity signal at those scales with the same fractions of Q and SF \mbox{``PrimB''}
galaxies (sub-samples Q$_{\rm 10}$ and SF$_{\rm 10}$) compared to the case separating between Q and SF galaxies resembles the higher assembly-type bias for the former than the latter.

We also separated the samples according to their stellar age because it has been proposed as an observational tracer for the halo formation time at fixed halo mass. We confirmed that the stellar age correlates with halo age in our model. For the ``PrimAll'' sample, older central galaxies tend to live in early-formed halos (see Fig. \ref{fig_SA_HaloAge} in the appendix).
We found that
the difference in clustering between the oldest and youngest galaxies in the fiducial case (``PrimAll'', 
top-right panel of Fig. \ref{fig_AB_GC_10perc})
is much stronger than between the most
quenched and star-forming central galaxies.

The clustering between the oldest and youngest galaxies drastically reduces by a factor of $\sim$4 between 1 and $\sim$10 $\mpc$
in the case ``PrimB''. 
However, the assembly-type bias becomes stronger at larger separations, reaching a clustering similar to the fiducial case at $r > 50~\mpc$.   
We quantified the impact of removing low-mass central galaxies near groups and clusters by randomly moving the position of massive halos by a fixed distance of 5 $\mpc$ from their original positions. The cross-correlation function between the ``PrimB'' sample and the massive halos with their original and random
positions becomes equal at separations larger than 10 $\mpc$ (Fig. \ref{fig_crossAB_PrimB_clusters}), which is the same scale with the increase in the assembly-type bias strength seen in the case ``PrimB''. Therefore, there is an excess of clustering due to the presence of massive halos at scales larger than 10 $\mpc$, which contributes to the signal.
Again, we repeat the measurement of the quenched fraction of neighboring galaxies
as a function of distance for this set of primary samples and show the results in the bottom-right panel of Fig.
\ref{fig_AB_GC_10perc}.
As can be seen, the signal of conformity around ``PrimAll'' and ``PrimB' central galaxies with old and young stellar populations is higher than for the selection of samples using sSFR, mirroring the relative amplitudes of assembly-type bias for these two sets of samples.

We checked that the \sag\ model is consistent with other models such as the {\sc SIMBA} simulation \citep{Dave+2019} in that, at the same halo mass, central galaxies with high stellar masses tend to live in early-formed halos \citep{Cui+2021NatAs}. However, in our model, the older central galaxies are typically hosted by early-formed halos, which is different from \cite{Cui+2021NatAs} or \cite{WangK+2023_Late-formedHaloes}, whose models show that early-formed halos prefer to host blue or star-forming galaxies. 
\cite{WangK+2023_Late-formedHaloes} show that central galaxies in late-formed halos have higher fractions of quiescence by 2 to 8 percent than their early-formed counterparts at halo masses higher than 10$^{12.5}$ \msun. This difference decreases for halos between 10$^{12}$ and 10$^{12.5}$ \msun, which is the lowest mass range explored in that work. Therefore, we do not expect a large discrepancy in the stellar populations of central galaxies hosted by low-mass halos in all these models. In the low-mass regime, \cite{WangK_PengY2025} show that star-forming central galaxies tend to be more efficient at converting baryons into stars than quiescent central galaxies in the L-GALAXIES model \citep{Henriques+2015}, while other models give similar efficiencies for these two populations. The stellar mass to halo mass ratios for both populations are usually within the error bars in all these models.
We checked that \sag\ is more similar to  {\sc IllustrisTNG300} in this regard because, in the same halo mass range, 10 percent of the most quiescent galaxies show a slightly higher stellar mass to halo mass ratio than the 10 percent of the most star-forming central galaxies, although within the error bars.

\cite{WangK+2023_Late-formedHaloes}
show that SDSS central galaxies in early-formed halos have higher fractions of star-forming galaxies. They use the residual $t_{\rm form}$ with respect to their median as a function of halo mass to define the halo formation time, where $t_{\rm form}$ is the lookback time when the halo has first assembled half of its final mass along the main branch. This definition is comparable to the $\delta_t$ dimensionless parameter introduced in \cite{LacernaPadilla2011}, where the residual with respect to the median formation time (or stellar age) is normalized by the dispersion around the median in units of time. 
\cite{Lacerna+2014} used this dimensionless parameter and found that older SDSS central galaxies hosted by low-mass halos exhibit higher clustering than younger galaxies at a fixed halo mass, as evidence of assembly-type bias. 

The low mass assembly-type bias can be explained by three causes, according to \cite{MansfieldKravtsov2020}: large-scale tidal fields, gravitational heating due to the collapse of large-scale structures, and splashback subhaloes located outside the virial radius. \cite{Palma_Lacerna+2025} showed that the splashback galaxies do not affect the conformity signal in the \sag~model. Therefore, it is likely that the correlation between conformity and low-mass assembly-type bias is mainly given by tidal forces and gravitational heating occurring in similar large-scale structures such as filaments. These two effects can 
truncate the mass accretion histories of low-mass halos \citep[e.g.,][]{MansfieldKravtsov2020}.

Notably, the assembly-type bias that we measured in this paper is strongest for the fiducial samples,
also showing the strongest conformity signal.  This result suggests that there could exist a direct relation between the amplitude of these two effects. If two-halo conformity
is confirmed with observational samples as shown, for example, by \cite{Ayromlou+2023} using SDSS \citep{SDSS-DR7} and photo-$z$ galaxies of the dark energy spectroscopic instrument \citep[DESI,][]{Dey+2019, Zou+2019} samples, it could be taken as an indication of assembly-type bias in the real Universe.

%============================= 
\begin{acknowledgements} 
The authors thank the anonymous referee for the revision that helped improve the presentation of this work.
We would like to thank Antonio Montero-Dorta, Sergio Contreras, Vladimir Avila-Reese, Raúl Angulo, Aldo Rodríguez-Puebla, Simon White, Doris Stoppacher, and Nelvy Choque-Challapa    for comments and discussions. 
We also thank Andr\'es Ruiz, Yamila Yaryura and Cristian Vega for their support in organizing and providing the \sag\ data used in
this work. 
NP acknowledges support from Agencia Nacional de Investigación Científica y Tecnológica through grants PICT-2021-0700 and PICT-2023-0002.
DP acknowledges the support through ANID-Subdirección de Capital Humano/Doctorado Nacional/2024/21241817. 
The CosmoSim database used in this paper is a service by the Leibniz-Institute for Astrophysics Potsdam (AIP). The MultiDark database was developed in cooperation with the Spanish MultiDark Consolider Project CSD2009-00064.

\end{acknowledgements}

\bibliographystyle{aa}
\bibliography{references}

%============================= 
\begin{appendix}

\section{Correlation between halo formation time and stellar age in \sag\ model}

\begin{figure}[hb!]
\includegraphics[width=0.5\textwidth]
{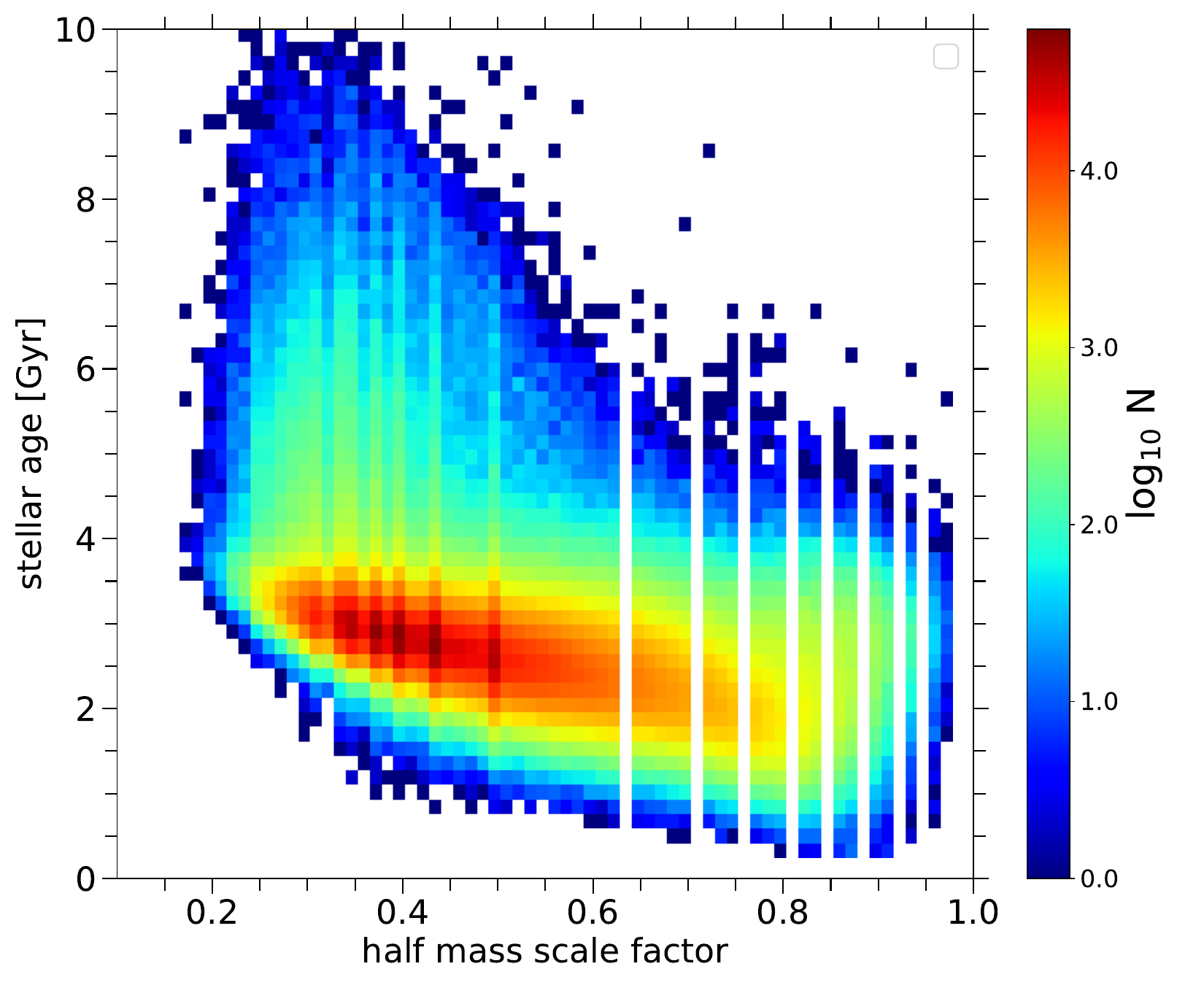}
\caption{Stellar age as a function of halo formation time. For the latter, we used the scale factor at which the most massive progenitor reaches half of the peak mass ($0.5 \times M_{peak}$) over accretion history. The number density of PrimAll galaxies is indicated in the color bar.
}
\label{fig_SA_HaloAge}
\end{figure}

The stellar age has been suggested as a tracer for the halo formation time \citep[e.g.,][]{LacernaPadilla2011}. Figure \ref{fig_SA_HaloAge} shows the correlation between stellar age for the ``PrimAll'' galaxies and the half-peak mass scale factor as a definition of halo formation time. For this definition of halo age, we used the parameter `halfmass\_scale' from the \textsc{\footnotesize{CosmoSim}} database that is the scale factor at which the most massive progenitor reaches half of the peak mass ($0.5 \times M_{peak}$) over accretion history.
The figure shows that in \sag\ older central galaxies, hosted by low-mass halos in the narrow range of 10$^{11.6}$ $\leq$ \mh/$\mhalo$ $<$ 10$^{11.8}$, tend to live in early-formed halos.

%%%fig stellar age vs halo age
\FloatBarrier

\clearpage
\end{appendix}
%%%%%%

\end{document}